# The basic reproduction number as a predictor for epidemic outbreaks in temporal networks

**Petter Holme**[1,2,3] **and Naoki Masuda**[4]

[1]Department of Energy Science, Sungkyunkwan University, 440-746 Suwon, Korea
[2]IceLab, Department of Physics, Umeå University, 90187 Umeå, Sweden
[3]Department of Sociology, Stockholm University, 10961 Stockholm, Sweden
[4]Department of Engineering Mathematics, University of Bristol, BS8 1UB, Bristol, UK

E-mail address: holme@skku.edu

## Abstract

The basic reproduction number $R_0$—the number of individuals directly infected by an infectious person in an otherwise susceptible population—is arguably the most widely used estimator of how severe an epidemic outbreak can be. This severity can be more directly measured as the fraction people infected once the outbreak is over, $\Omega$. In traditional mathematical epidemiology and common formulations of static network epidemiology, there is a deterministic relationship between $R_0$ and $\Omega$. However, if one considers disease spreading on a temporal contact network—where one knows when contacts happen, not only between whom—then larger $R_0$ does not necessarily imply larger $\Omega$. In this paper, we numerically investigate the relationship between $R_0$ and $\Omega$ for a set of empirical temporal networks of human contacts. Among 31 explanatory descriptors of temporal network structure, we identify those that make $R_0$ an imperfect predictor of $\Omega$. We find that descriptors related to both temporal and topological aspects affect the relationship between $R_0$ and $\Omega$, but in different ways.

## Introduction

The interaction between medical and theoretical epidemiology of infectious diseases is probably not as strong as it should. Many results in the respective fields fail to migrate to the other. There are of course exceptions. Perhaps the most important are the ideas of epidemic thresholds and the parameter $R_0$—the basic reproduction number—as a key predictor of the epidemiological severity of a disease [1,2]. $R_0$ is defined as the expected number of others that an infected individual will infect if he or she enters a population entirely composed of susceptible individuals. It is thus a combined property of the process of contagion and the contact patterns of the population. In classic mathematical models of infectious disease spreading, $R_0 = 1$ marks an epidemic threshold. If $R_0 < 1$, the expected total number of infected people in an outbreak, denoted by $\Omega$, will not depend on the total population size $N$. If $R_0 > 1$, the expected value of $\Omega$ is proportional to $N$. In other words, in the limit of large populations, a finite fraction of the population can be infected. The focus on $R_0$ in the literature has sometimes been so strong that researchers rather calculate $R_0$ than quantities directly related to the outbreak, such as prevalence, incidence, and time to the peak prevalence.

The use of $R_0$ is not entirely unproblematic. First, it is hard to estimate both in models [3–5] and from outbreak data [6–8]. Second, the result that $R_0 = 1$ defines an epidemic threshold rests on very coarse assumptions [3,9,10].

For example, one needs to assume that every pair of individuals has the same chance of interacting at any given time. In fact, interaction rates depend on pairs of individuals—people living in the same city are more likely to interact than those living in different cities. The derivation of $R_0$ has been extended to the case in which information about contact networks (describing who can spread the disease to whom) is available [11–15]. In this case, the derivation is usually restricted to the case of regular networks, where all individuals have the same degree (number of neighbors in the contact network) [14,15]. Sometimes people use definitions of $R_0$ that differs from the original [11–13,16] in a strict sense (but typically captures some similar property relevant for the modeling framework in question). The assumption that a pair of individuals interacts at the same rate over time does not hold true in reality either. For example, interaction is more likely to take place when most people are awake. This point is reason for the increasing interest in temporal networks (showing who is in contact with whom, at what time) as a representation for the interactions underlying epidemic spreading, which focus on time dependence of networks [17–19]. There have been a few attempts to examine $R_0$ for temporal networks. Ref. [16,20], for example, derives $R_0$ for a specific model of temporal networks. Ref. [21] measures $R_0$ in empirical temporal networks, but does not relate it to prevalence, final outbreak size or other direct measures of outbreak severity.

One possible approach to $R_0$ is to improve it—to find more accurate estimators of disease severity. However, $R_0$ is routinely estimated for different infectious diseases by public health organizations worldwide. These estimates constitute an important resource for monitoring and comparing disease outbreaks. Rather than discarding this data by proposing another quantity, we will investigate what $R_0$ really tells us about disease spreading in empirical temporal networks of human contacts. Including the temporal information can make a big impact on the outbreak dynamics compared to modeling epidemics on a static network, let alone a fully mixed model [17–19].

We use the Susceptible–Infectious–Recovered (SIR) model with constant disease duration [22]. This model has two control parameters—the probability of disease transmission (upon a contact between an infectious and susceptible individual), denoted by λ, and the duration of the infectious stage, denoted by δ. We numerically simulate the SIR model on various temporal networks. First, we observe that in this case Ω is not uniquely determined from an $R_0$ value. A combination of λ and δ can give a larger $R_0$ but a smaller Ω than another combination does. Then, we investigate how the structure of the temporal contact network explains the relationship between $R_0$ and Ω. Instead of building a theory that bridges the microscopic structure of temporal network data and the emergent properties of the outbreak, we screen many potentially interesting descriptors of the temporal network structure by identifying those that are strongly correlated with the descriptors of the shape of scatter plots of Ω vs. $R_0$.

## Results

**Empirical data**

We analyze empirical sequences of contacts between people. These data sets can be divided into physical proximity and electronic communication data. The former type could be interesting for studying information and disease spreading mediated by human contacts. The latter type is primarily of interest in the context of information spreading (bearing in mind that information spreading not necessarily follows the same dynamics as infectious diseases). In all data sets, nodes are human individuals. We list some basic statistics of the data sets in Table 1.

One data set belonging to the physical proximity class comes from the *Reality mining* study [23], where contacts between university students were recorded when their smartphones were within Bluetooth range (10~15 m). We use the same subset of this data as in Ref. [24]. Another class of proximity data was collected from groups of people wearing radio-frequency identification sensors. One such dataset comes from the attendees of a conference [25] (*Conference*), another from a school (*School*) [26], another from a hospital (*Hospital*) [27] and yet another from visitors to a gallery (*Gallery*) [25]. *School* and *Gallery* are collected for two and 69 days, respectively. We analyze the days separately and average the results over the days. In these data sets, a contact between people closer than 1~1.5m was recorded every 20 seconds. Finally, we use a data set of sexual contacts between sex sellers and buyers collected from a Brazilian web forum (*Prostitution*) [28].

The class of electronic communication data includes two e-mail networks. These data sets are described in detail in Refs. [29] (*E-mail 1*) and [30] (*E-mail 2*). E-mails have a natural direction from the sender to the recipient. However, to analyze all the data sets in the same way, we treat them as undirected temporal networks. We furthermore study two Internet communities: a dating community (*Dating*) [31] and a film community (*Online community*) [32]. The contacts in these data sets represent messages from person to person like e-mails do. In *Dating* there are also "flirts" with which one user expresses interest in another (but does not send text, images or other information). A slightly different form of online pair-wise interaction is posting to public web pages. We study one data set of posts to the home page ("wall") of *Facebook* [33] and a data set from the aforementioned film community where a contact represents a reply to a post at a public forum (*Forum*) [32]. One contact in these data sets is thus a publically accessible message from one user to another.

**Final outbreak size as a function of $R_0$**

In Fig. 1, we show scatter plots of $\Omega$ vs. $R_0$ for our data sets. One scatter plot corresponds to one data set. More precisely, we measure $R_0$ directly from the simulations according to the definition—the average number of others infected by the infection source. $\Omega$ is the fraction of recovered individuals when the outbreak has subsided, i.e., when there no longer are infectious individuals. A point in a scatter plot represents an average over $10^6$ runs for given parameter values $(\lambda,\delta)$. Each run starts with one infected node that is selected from all nodes with the equal probability. We assume the source of the infection is infected at the time of the first contact. In total, we sample 20×20 points in the $(\lambda,\delta)$ parameter space, where each parameter varies from 0.001 to 1 with exponentially increasing intervals. $\delta$ is defined as a fraction of the total sampling time.

For all the data sets, there is a significant deviation from a deterministic relationship between $R_0$ and $\Omega$. Here, a deterministic relationship is operationally defined as the situation in which the $\Omega$ value is uniquely determined by the value of $R_0$ (as it would be in most fully mixed and network models we are aware of). Interestingly, the way these scatter plots deviate from a deterministic relationship depends on data sets. For example, for the *Hospital* data red points are typically on top of the green ones—i.e. points with higher $\lambda$ and lower $\delta$ give larger outbreaks than points with similar $R_0$ but lower $\lambda$ and higher $\delta$. For the *Facebook* data the situation is reversed.

**Characterizing the shape of the $\Omega$ vs. $R_0$ point cloud**

To explore the causes of the imperfectness of $R_0$ as a predictor of $\Omega$, we define six so-called shape descriptors, which measure the shape of the point clouds shown in Fig. 1. The shape descriptors are listed in Table 2, their definitions are illustrated in Fig. 2.

The first shape descriptors is the Kendall's $\tau$ (Fig. 2A), which captures how good $R_0$ is as a predictor of $\Omega$. We chose Kendall's $\tau$ because the $\Omega$ vs. $R_0$ curve is highly non-linear such that the Pearson's correlation coefficient would underestimate how good a predictor $R_0$ is. Among non-linear correlation measures, Kendall's $\tau$, is the most principled and easiest to understand. It counts the number of point pairs that are connected by a line with a positive slope (*concordant pairs*) and a negative slope (*discordant pairs*). Kendall's τ is then the number of concordant pairs minus the number of discordant pairs divided by the total number of pairs. In the context of measuring the $R_0$-$\Omega$ correlation, we denote Kendall's τ by $\tau_{R_0\Omega}$.

Next four shape descriptors focus on the region in the ($R_0$,$\Omega$) space where the spread of the points is the largest (Fig. 2B, C). We look for the discordant (λ,δ) pair with the largest difference between the its $R_0$ values. This difference defines $\Delta_{R_0}$. Similarly, the largest difference in $\Omega$ among discordant pairs defines $\Delta_\Omega$. We also measure the average $R_0$ value, $\rho_{R_0}$, of the two $R_0$ values derived from the discordant pair maximally separated in $R_0$. Similarly $\rho_\Omega$ is the average $R_0$ value of the discordant pair maximally separated in $\Omega$. The shape descriptors $\rho_{R_0}$ and $\rho_\Omega$ thus show the locations on the $R_0$ axis of the maximally separated discordant pairs. They may be related to the location of the epidemic threshold, where $\Omega$ takes off from zero in an infinite population.

As mentioned above, for some data sets, given a value of $R_0$, higher δ implies higher $\Omega$ (*Hospital*), whereas the relationship is reversed for other data sets. To quantify this observation, we define the sixth shape descriptor $\tau_{\alpha\Omega}$ that we call *λδ-balance* for short. To define $\tau_{\alpha\Omega}$, we start by dividing the range of $R_0$ into ten equidistant bins between the smallest and largest observed values (Fig. 2D). Within a bin, the points have fairly similar $R_0$ values, but their λ and δ values can be diverse. To measure the effect of the balance between λ and δ on $\Omega$, we calculate the angle α that a (λ,δ) pair relative to the origin makes to the diagonal in the (λ,δ)-plane, i.e., the λ = δ line (Fig. 2E). Then, we measure the correlation between α and $\Omega$ by Kendall's τ (Fig. 2F). Finally, we average the values for the different bins. To avoid confusion, we denote the calculated Kendall's τ by $\tau_{\alpha\Omega}$.

**Temporal and static network descriptors**

To characterize the structure of the contact structures modeled as temporal networks, we use 31 different quantities, which we call network descriptors. They are listed in Table 3. We have chosen quantities that are relatively simple and intuitive.

*Time evolution*

We calculate eight network descriptors that characterize the long-term behavior of the contact dynamics—basically, how the contacts process differs from a stationary process. The background is that some of these data sets (e.g. *Prostitution*, *Dating*, *Forum* and *Online* community) are growing throughout the sampling period. A fast-spreading outbreak would thus, effectively, spread in a larger population (defined as the set of individuals possible to be infected) in the end than in the beginning. The *Gallery* data is also special in that the individuals in the beginning of the sampling are not present in the end. Ref. [34] argues, in more general terms, that when the first and last contacts of a link (pairs of nodes that are in contact at least once) happen is important for the behavior of outbreaks.

The first such set of quantities focuses on the time when nodes and links appear for the first time. For example, Ref. [34] points at the growth of the *Prostitution* data set as a factor behind the observation [35] that the order of events speeds up disease spreading in this data. We use *f* to symbolize this class of network descriptors. We measure the fraction of links present at half the sampling time relative to the final number of links. Because several studies in temporal networks address the role of the order of events [35,36], rather than the time itself, we also measure the corresponding quantities if time is replaced by the contact index (the index of the contact number—1 for the first contact, 2 for the second, etc.). These have subscript 'C' as opposed to 'T' for time. Furthermore, the descriptors concerning nodes and links have the subscripts 'N' and 'L', respectively.

Another class of network descriptors, denoted $F$, focuses on persistent nodes or links. $F$ is the fraction of nodes (subscript N) or links (subscript L) present in the first and last 5% of time (T) or contact index (C). Figure 3 illustrates $f$ and $F$.

These network descriptors calculated across the different data sets span a relatively wide range. For example, $f_{NT}$, the fraction of links present at half the sampling time, takes values from 0.17 (*Facebook*) to 0.98 (*School*).

*Node and link activity*

The node activity descriptors relate to the bursty nature of human activity as characterized by intense periods of activity separated by long periods of quiescence [37]. To characterize burstiness, one usually starts from *interevent times*, i.e., the times between consecutive contacts for a node or link. For simplicity, we ignore correlations between consecutive interevent times and focus on the probability distribution of interevent times. The distribution is often right-skewed—a structure that has been shown to slow down epidemic spreading [38–41]. To characterize the distribution, we measure four descriptors, i.e., the mean μ, standard deviation σ, coefficient of variation $c$ (i.e. the standard deviation divided by the mean) [37], and the sample skewness given by

$$\gamma = \frac{\sqrt{n(n-1)}}{n-2} \frac{\mu_3}{\mu_2^{3/2}} \qquad (1)$$

where μ$_2$ and μ$_3$ are the second and third moments of the distribution, respectively.

Some studies have pointed out that the duration of presence of a node or link in the data can be more important for spreading dynamics than interevent times [34,42]. For this reason, we also study the distribution of node and link durations and use the same four descriptors. In sum, we use 16 network descriptors in this category—μ, σ, $c$ and γ for interevent times and duration of activity, for both nodes and links.

*Degree distribution*

In the following, we define static network descriptors, i.e., those for aggregate contact networks. Among them, the degree distribution is arguably the most important for disease spreading. A right-skewed degree distribution, which is observed in many empirical networks, is known to facilitate disease spreading [43]. For simplicity, we use the network of accumulated contacts (even though one may be able to find network representations of temporal network data that better captures the important structures for disease spreading [44]). To summarize the shape of the degree distribution, we use the same four descriptors as for the interevent time and duration distributions—μ, σ, $c$ and γ.

*Other static network descriptors*

We also measure other static network descriptors. First, we count the number of nodes, $N$. Because the number of links is equal to the half of the mean degree times $N$, we do not include it in the analysis.

We also measure the degree assortativity $r$ (essentially, the Pearson correlation coefficient of the degrees at either side of a link). This network descriptor measures the tendency for assortative mixing by degree, i.e., whether high-degree nodes tend to connect to high-degree nodes and low-degree nodes to low-degree nodes. It has been shown that assortativity affects disease spreading (exactly how depends on the specific epidemic model and other structures of the contacts) [45–48].

Finally, we measure the clustering coefficient—the number of triangles in the network divided by the number of connected triples (not necessarily a full triangle) normalized to the interval [0,1]. Similar to assortativity, the relative number of triangles (clustering) is also a contact-structural factor influencing disease dynamics [46–51]. As an example, if we compare SI disease spreading on a clustered network with a random network with the same number of nodes and links, the early stage of the spreading would be faster in the less clustered network [49,50]. Intuitively, if a disease spreads from one individual to two neighbors, and the three individuals are connected as a triangle, then the third link of the triangle is useless for the spreading process. If the third link were connected elsewhere, the disease would spread faster.

**Structural determinants of the $\Omega$ vs. $R_0$ point cloud**

Ultimately, one would like to explain how the relations between $R_0$, $\Omega$, $\lambda$ and $\delta$ emerge from the contact structure. In this work, as mentioned, we take a different approach and look at the Pearson correlation coefficient between the shape descriptors (Table 2) and network descriptors (Table 3). In this way, we search for network descriptors that contribute to the deviation from a deterministic relationship between $\Omega$ and $R_0$. A temporal network data set defines a data point that is fed to the calculation of the correlation coefficient; there are 12 data points available for regression analysis. We decided to use the Pearson correlation coefficient and not multivariate regression methods because there are 31 dependent variables, i.e., network descriptors (and 6 independent variables, i.e., shape descriptors), whereas we have only 12 data points.

In Fig. 4, we plot the results from our correlation analysis. In each panel, we plot the coefficient of determination $R^2$ (square of the Pearson correlation coefficient) between a shape descriptor and each network descriptor. The network descriptors are grouped in accordance with the subsections of the previous section.

The predictability of $R_0$ with respect to $\Omega$, as measured by $\tau_{R_0\Omega}$ (Fig. 4A), is to some extent ($p < 0.05$) explained by the coefficients of variation of the interevent time for the node and link interevent time distribution, $c_{Lt}$ and $c_{Nt}$. This correlation is positive, so broader interevent time distributions (burstier contact patterns) imply worse predictability. Furthermore, the α dependence of $\Omega$ is most strongly correlated with the burstiness of the nodes $c_{Nt}$. In this case the correlation is negative. This means that if we compare two points with the same $R_0$ value, where the first parameter set has a comparatively large transmission probability and short disease duration than the second, then the first parameter set tends to trigger a larger outbreak size than the second. These quantities are strongly affected by burstiness. The remaining four shape descriptors concern the location (in the ($R_0$,$\Omega$) space) of the biggest deviation from a deterministic relationship and the size of the deviation.

Figure 4C shows the correlation coefficient with the location along the $R_0$ axis of the mid-point of the discordant pair with the largest separation in $R_0$, i.e., $\rho_{R_0}$. Also in this case, network descriptors derived from the interevent-time distributions are relatively strongly correlated with $\rho_{R_0}$. The mean $\mu_{NT}$ and standard deviation $\sigma_{NT}$ as well as the skewness $\gamma_{NT}$ show strong correlations. Furthermore, the fraction of links present in both the first and last 5% of the contacts ($F_{LC}$) shows an $R^2 = 0.4$ correlation with $\rho_{R_0}$ ($p = 0.06$). Furthermore, even though they do not reach the $p < 0.05$ significance criterion, other link-related quantities of the time evolution ($\mu_{Lt}$, $c_{Lt}$, $\gamma_{Lt}$, $\mu_{L\tau}$, $\sigma_{L\tau}$, $c_{L\tau}$ and $\gamma_{L\tau}$) show $R^2$ values over 0.3. Figure 4D indicates that the largest width of a discordant pair, $\Delta_{R_0}$, is strongly correlated with a number of temporal network descriptors. First, $\Delta_{R_0}$ is correlated with both those relating to the node and link activity when the real time, not the contact index, is used ($\mu_{Lt}$, $c_{Lt}$, $\gamma_{Lt}$, $\mu_{Lt}$, $\sigma_{Lt}$ and $\gamma_{Lt}$). Second, $\Delta_{R_0}$ is correlated with the time evolution, especially with the $F$ quantities—measuring the fraction of links and nodes present both in the beginning and end of the sampling period ($f_{NC}$, $f_{NT}$, $F_{NC}$, $F_{LC}$, $F_{NT}$, $F_{LT}$); $p < 0.01$). Figure 4E shows the correlation with the $R_0$-location with the discordant pair with the largest separation in $\Omega$, $\rho_\Omega$. Just like $\rho_{R_0}$ (Fig. 4C), much of the variance in $\rho_\Omega$ is explained by the time-related descriptors in real time ($f_{LC}$, $f_{LT}$, $F_{LC}$, $F_{LT}$ and $\sigma_{Lt}$). More interestingly, the largest $\Omega$-separation of discordant pairs, $\Delta_\Omega$ (Fig. 4F) is strongly and positively correlated with some static network descriptors, i.e., the coefficient of variation and the skewness of the degree distribution ($c_k$ and $\gamma_k$).

## Discussion

In this work, we have shown that temporal network structure of human contacts can change the interpretation of the basic reproduction number $R_0$. We have found pairs of SIR parameter values $(\lambda_1, \delta_1)$ and $(\lambda_2, \delta_2)$ such that $R_0(\lambda_1, \delta_1) < R_0(\lambda_2, \delta_2)$ and $\Omega(\lambda_1, \delta_1) > \Omega(\lambda_2, \delta_2)$. In other words, the expected number of secondary infections of the outbreak's source is smaller for $(\lambda_1, \delta_1)$ than $(\lambda_2, \delta_2)$, but the expected final fraction of individuals that had the infection is larger for $(\lambda_1, \delta_1)$ than $(\lambda_2, \delta_2)$. It is hard to give a succinct explanation for this phenomenon, and we do not attempt that in the present paper. It relates to many aspects of the contact patterns—static network structures, dynamic network structures, and the fact that empirical data is finite-sized, non-equilibrium and inhomogeneous [18,19,52]. On the other hand, it is easy to imagine scenarios where this happens. Assume, for simplicity, that $\lambda_1 \ll \lambda_2$, $\delta_1 \gg \delta_2$ and the nodes split in two halves—one half active throughout the sampling time, the other half entering after some time. Then, in the $(\lambda_2, \delta_2)$ scenario, the larger $\lambda$ (i.e., $\lambda_2$) could cause a burnout outbreak that ends before the second group of nodes enters the system. Therefore, $R_0$ would be high, whereas $\Omega$ does not exceed 1/2. In the $(\lambda_1, \delta_1)$ scenario, $R_0$ would be smaller. However, the duration of infection would be long enough for the second half of the nodes to be infected, so $\Omega$ could be larger than 1/2. Therefore, a larger value of $R_0$ does not necessarily mean that the disease spreads more easily. At the same time, the correlation between $R_0$ and $\Omega$ is often strong, especially if one accepts a non-linear relationship. For most practical purposes, it probably suffices to assume that $R_0$ is a good predictor of $\Omega$.

Looking closer at the deviation of the $\Omega$ vs. $R_0$ scatter plots from a deterministic relationship and structural correlates of the amount of the deviation, we notice that a combination of seemingly unrelated descriptors of temporal network structure often shows a significant correlation. This result suggests that—although a better achievement may be obtained through identification of microscopic factors contributing to these phenomena—such factors could be interdependent and hard to fully disentangle. Probably a fruitful path would be to vary the structure in models of contact patterns and look at responses in the $\Omega$ vs. $R_0$ plots. However, already based on the current numerical results, we can draw some conclusions. One of them is that the temporal network factors often seem important. In particular, the quantities relating to the interevent-time distributions are significant predictors of e.g. the overall correlation between $\Omega$ and $R_0$. This is a bit surprising in the light of Refs. [35] and [41] that have found that the birth and death of links and nodes influence (some other quantities relating to) spreading phenomena (probably also the importance of the "loyalty" metrics in Ref. [52]). Only one aspect of the $\Omega$ vs. $R_0$ plots—$\Delta_\Omega$ (see Table 2 and Fig. 2C for definition)—is primarily explained by the static network properties, specifically the coefficient of variation and skewness of the degree distribution. This result is accompanied by the largest confidence level ($p < 0.001$) of the correlation. In contrast to $\Delta_\Omega$, a similar shape descriptor $\Delta_{R_0}$ (see Table 2 and Fig. 2B for definition) is strongly correlated with several of the temporal network properties and not with the static ones. Especially the former observation is interesting—even though temporal structure is needed to see any spread in $\Delta_\Omega$ at all, it is the degree distribution that is the most strongly correlated with the actual value of $\Delta_\Omega$.

Needless to say, this work opens more questions than it answers. In particular, it calls for mechanistic modeling connecting $R_0$ and $\Omega$. Another direction would be to develop improved estimators of disease severity.

## Methods

In this section, we will go through technicalities of the methods that are not fully explained in the Results section.

### SIR simulations

In this work we use the constant duration SIR model. We initialize all individuals to susceptible and pick one random individual $i$ to be the source of the infection. We assume that $i$ becomes infected at the same time as its first appearance in the data. In a contact between an infectious and susceptible, the susceptible will (instantaneously) become infectious with a probability $\lambda$. Infectious individuals stay infectious for $\delta$ time steps after which they become recovered. If many contacts happen during the same time step, we go through them in a random order.

A more common version of the SIR model is to let infectious individuals recover with a constant rate. Qualitatively, both versions give the same results [21]. We use the constant duration version because it is a bit more realistic [53,54] and makes the code a bit faster than the exponentially distributed durations.

### Measuring the $\lambda\delta$-balance

A combination of a large $\lambda$ and small $\delta$ can give the same $R_0$ value as a combination of a small $\lambda$ and large $\delta$. At the same time, $\Omega$ may depend on one of these parameters more strongly than on the other. The result is a vertical trend in the colors of the points as seen in Fig. 1 (most clearly for the *Forum*, *Dating* and *Online community* data). We measure this tendency—the $\lambda\delta$-balance—as illustrated in Fig. 2D, E, and F. First, we segment the $R_0$ axis into ten bins. The number of bins is determined based on a trade-off between minimizing the spread of the points along the $R_0$ axis, and maximizing the number of points per bin. After the division into bins, we capture the $\lambda\delta$-balance via the angle $\alpha$ between the line from the origin to the parameter value $(\lambda,\delta)$ and the $\lambda = \delta$ line. Finally, we calculate Kendall's $\tau$ for the relationship between $\alpha$ and $\Omega$ and average the $\tau$ values over all bins.

## Acknowledgment

PH acknowledges support from Basic Science Research Program through the National Research Foundation of Korea (NRF) funded by the Ministry of Education (2013R1A1A2011947) and the Swedish Research Council. NM acknowledges support from JST, CREST, and JST, ERATO, Kawarabayashi Large Graph Project.

# Tables

|  | Number of individuals | Number of contacts | Sampling time | Time resolution |
|---:|---:|---:|---:|---:|
| *Conference* | 113 | 20,818 | 2.5d | 20s |
| *Dating* | 28,972 | 529,890 | 512.0d | 1s |
| *E-mail 1* | 57,189 | 444,160 | 112.0d | 1s |
| *E-mail 2* | 3,188 | 115,684 | 81.6d | 1s |
| *Facebook* | 293,878 | 876,993 | 4.36y | 1s |
| *Forum* | 7,084 | 1,412,401 | 8.61y | 1s |
| *Gallery* | 159(8) | 6,027(350) | 7.32(11)h | 20s |
| *Hospital* | 293,878 | 64,625,283 | 9.77y | 1d |
| *Online community* | 35,624 | 472,496 | 8.27y | 1s |
| *Prostitution* | 16,730 | 50,632 | 6.00y | 1d |
| *Reality mining* | 64 | 26,260 | 8.63h | 5s |
| *School* | 237(1) | 62,886(2,263) | 8.61(3)h | 20s |

**Table 1.** The basic statistics of the data sets. The numbers in parenthesis indicate the standard deviation in order of the last digit for the two composite data sets (*Gallery* and *School*).

| Symbol | Definition | Explained in |
|---|---|---|
| $\tau_{R_0\Omega}$ | Kendall's τ of $R_0$ vs. Ω (number of concordant pairs of parameter values – number of discordant pairs) / total number of pairs | Fig. 2A |
| $\tau_{\alpha\Omega}$ | λδ-balance, the Kendall's τ of α vs. Ω averaged over ten equal sized bins of $R_0$. α is the angle to the λ = δ line of a point in λ,δ-space | Fig. 2D–F |
| $\Delta_{R_0}$ | Largest difference in $R_0$ among discordant pairs of points in $R_0$-Ω space, where one point corresponds to one combination of λ and δ. | Fig. 2B |
| $\Delta_\Omega$ | Largest difference in Ω among discordant pairs of points in $R_0$-Ω space | Fig. 2C |
| $\rho_{R_0}$ | Midpoint of the $R_0$ values of the pair defining $\Delta_{R_0}$ | Fig. 2B |
| $\rho_\Omega$ | Midpoint of the $R_0$ values of the pair defining $\Delta_\Omega$ | Fig. 2C |

**Table 2.** Shape descriptors for the point clouds shown in Fig. 1.

| Symbol | Definition |
| --- | --- |
| $f_{NC}$ | Fraction of nodes present (i.e. having had at least one contact) when half of the contacts happened. |
| $f_{NT}$ | Fraction of nodes present at half the sampling time. |
| $f_{LC}$ | Fraction of links present when half of the contacts happened. This is illustrated in Fig. 3A and B. |
| $f_{LT}$ | Fraction of links present at half the sampling time. |
| $F_{NC}$ | Fraction of nodes present at both the first and last 5% of the contacts. |
| $F_{NT}$ | Fraction of nodes present at both the first and last 5% of the sampling time. This is illustrated in Fig. 3C and D. |
| $F_{LC}$ | Fraction of links present at both the first and last 5% of the contacts. |
| $F_{LT}$ | Fraction of links present at both the first and last 5% of the sampling time. |
| $\mu_{Lt}$ | Mean of interevent times over links. |
| $\sigma_{Lt}$ | Standard deviation of interevent times over links. |
| $c_{Lt}$ | Coefficient of variation of interevent times over links. In the terminology of Ref. [31], this is the burstiness of link activity. |
| $\gamma_{Lt}$ | Skewness of interevent times over links. |
| $\mu_{L\tau}$ | Mean of the number of other contacts between two consecutive contacts of a link. |
| $\sigma_{L\tau}$ | Standard deviation of the distribution of the number of other contacts in the data between two consecutive contacts of a link. |
| $c_{L\tau}$ | Coefficient variation of the distribution of the number of other contacts in the data between two consecutive contacts of a link. |
| $\gamma_{L\tau}$ | Skewness of the distribution of the number of other contacts in the data between two consecutive contacts of a link. |
| $\mu_{Nt}$ | Like $\mu_{Lt}$ but for nodes. |
| $\sigma_{Nt}$ | Like $\sigma_{Lt}$ but for nodes. |
| $c_{Nt}$ | Like $c_{Lt}$ but for nodes, i.e., the burstiness of node activity. |
| $\gamma_{Nt}$ | Like $\gamma_{Lt}$ but for nodes. |
| $\mu_{N\tau}$ | Like $\mu_{L\tau}$ but for nodes. |
| $\sigma_{N\tau}$ | Like $\sigma_{L\tau}$ but for nodes. |
| $c_{N\tau}$ | Like $c_{L\tau}$ but for nodes. |
| $\gamma_{N\tau}$ | Like $\gamma_{L\tau}$ but for nodes. |
| $\mu_k$ | Average degree of the network of accumulated contacts. |
| $\sigma_k$ | Standard deviation of the degree distribution of the network of accumulated contacts. |
| $c_k$ | Coefficient of variation of the degree distribution of the network of accumulated contacts. |
| $\gamma_k$ | Skewness of the degree distribution of the network of accumulated contacts. |
| $N$ | Number of nodes. |
| $C$ | Clustering coefficient of the network of accumulated contacts. |
| *r* | Degree assortativity of the network of accumulated contacts. |

**Table 3.** Descriptors of temporal network structure.

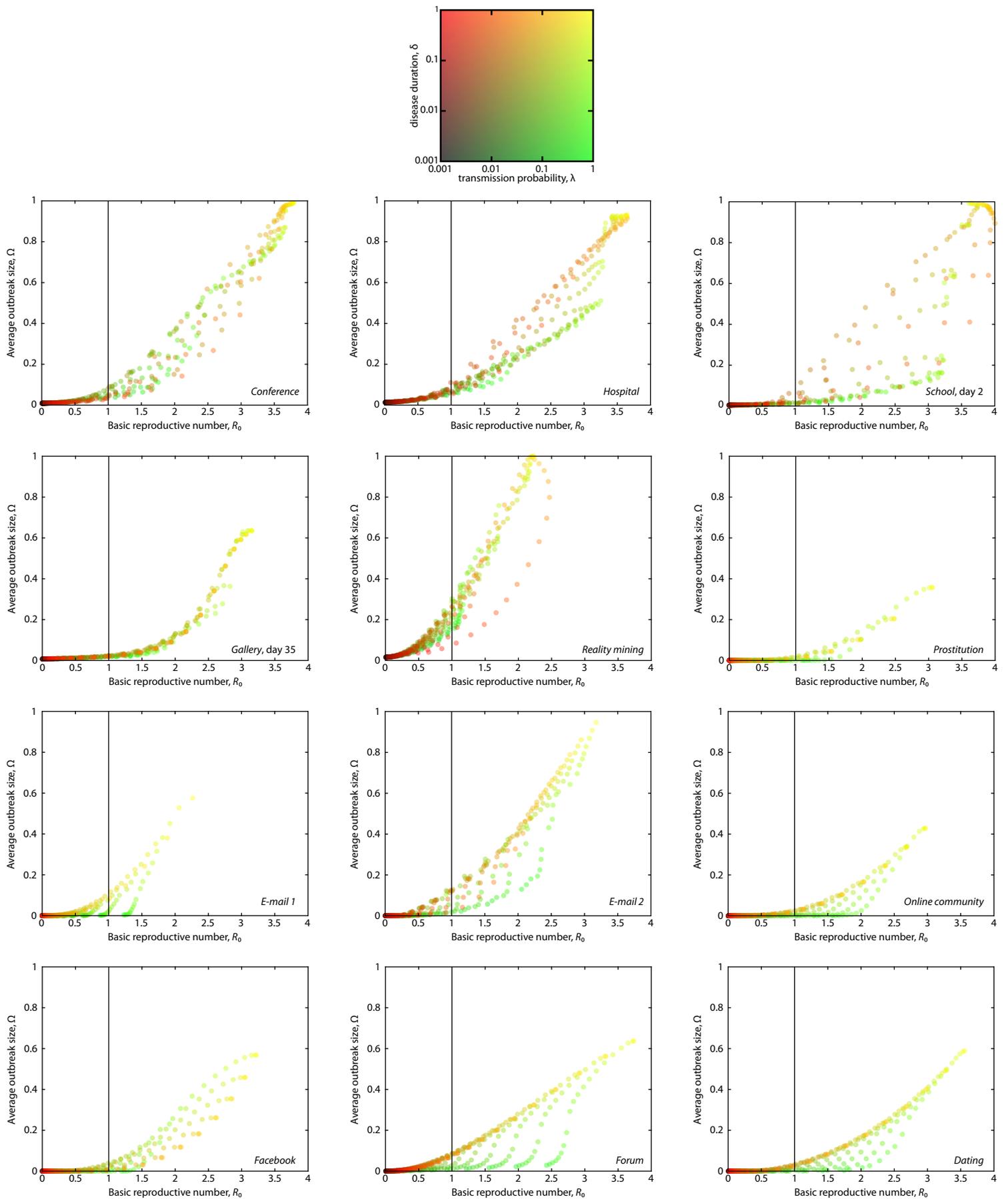

**Fig. 1. The average outbreak size plotted against the basic reproduction number for 12 data sets (indicated in the figure) of human interaction.** Each point of the scatter plots corresponds to one pair (λ,δ), where λ is the infection probability and δ is the duration of infection. In the upper left corner there is a legend for the color-coding of these points. In the other panels, a data point is an average over $10^4$ runs of the SIR model as described in the Methods section. The vertical lines mark $R_0 = 1$—the epidemic threshold for the canonical, fully mixed SIR model.

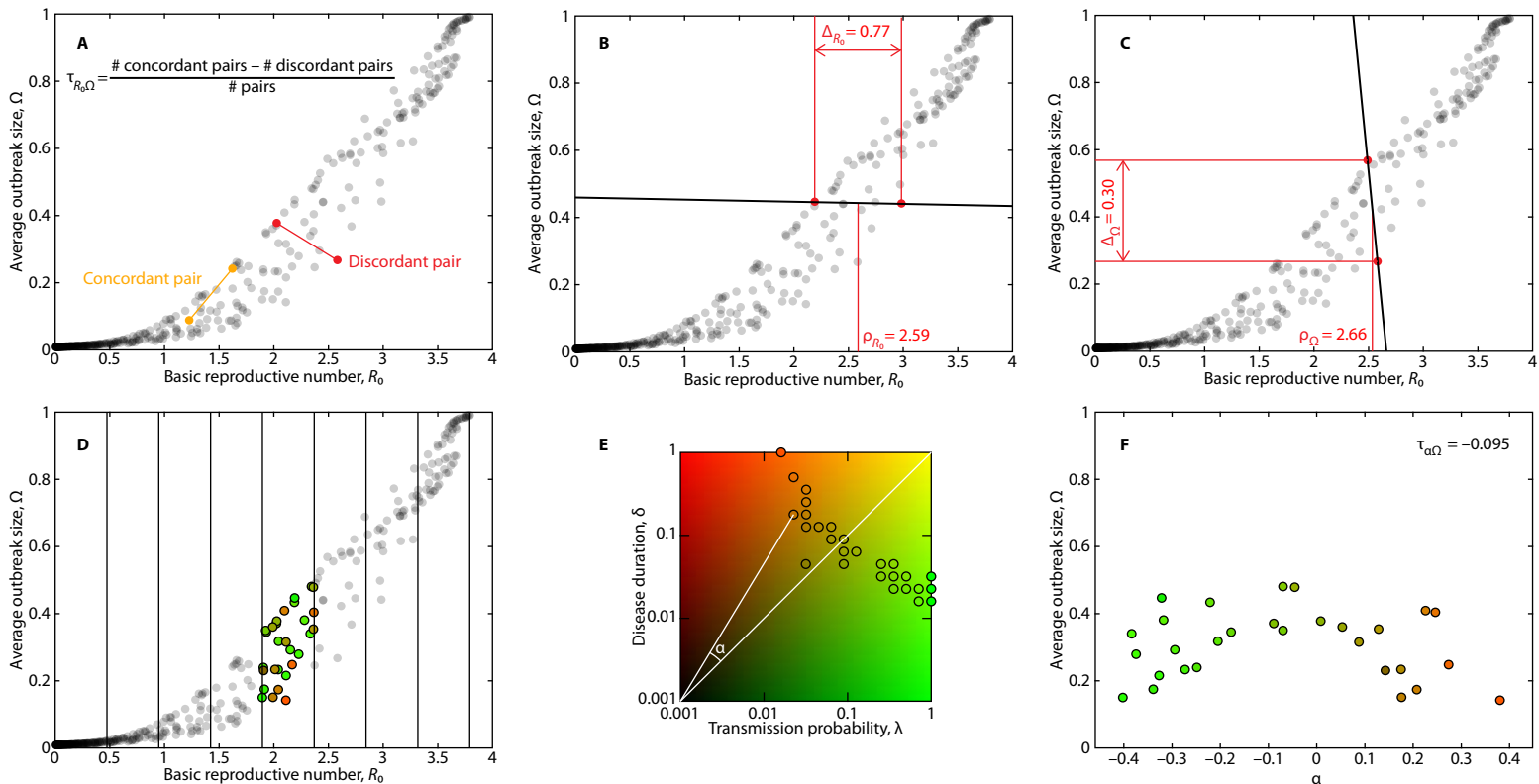

**Fig. 2. Explanation of shape descriptors to characterize the point clouds shown in Fig. 1.** All examples come from the Conference data set. Panel A describes Kendall's τ—a correlation coefficient based on the counting of discordant pairs (pairs of points connected by a line of negative slope). Panels B and C show the maximal separation of discordant pairs. In B, the measures focus on the pair with the largest separation in the $R_0$ direction. $\Delta_{R_0}$ denotes the maximum separation; $\rho_{R_0}$ is the mean $R_0$ value for the maximally discordant pair. Panel C shows the similar quantities, $\Delta_\Omega$ and $\rho_\Omega$, defined along the Ω direction. Panels D, E, and F illustrate the measurement of λδ-balance via $\tau_{\alpha\Omega}$. This descriptor captures the tendency of some data sets to have high-λ, low-δ points above high-δ, low-λ points, while for other data sets, the situation is reversed. Panel D illustrates how the $R_0$ axis is segmented into bins. Panel E shows how we assign a (λ,δ)-plane angle, α, to all points in the bin. Panel F shows how we measure the correlation between α and Ω, which is very weak in this particular case.

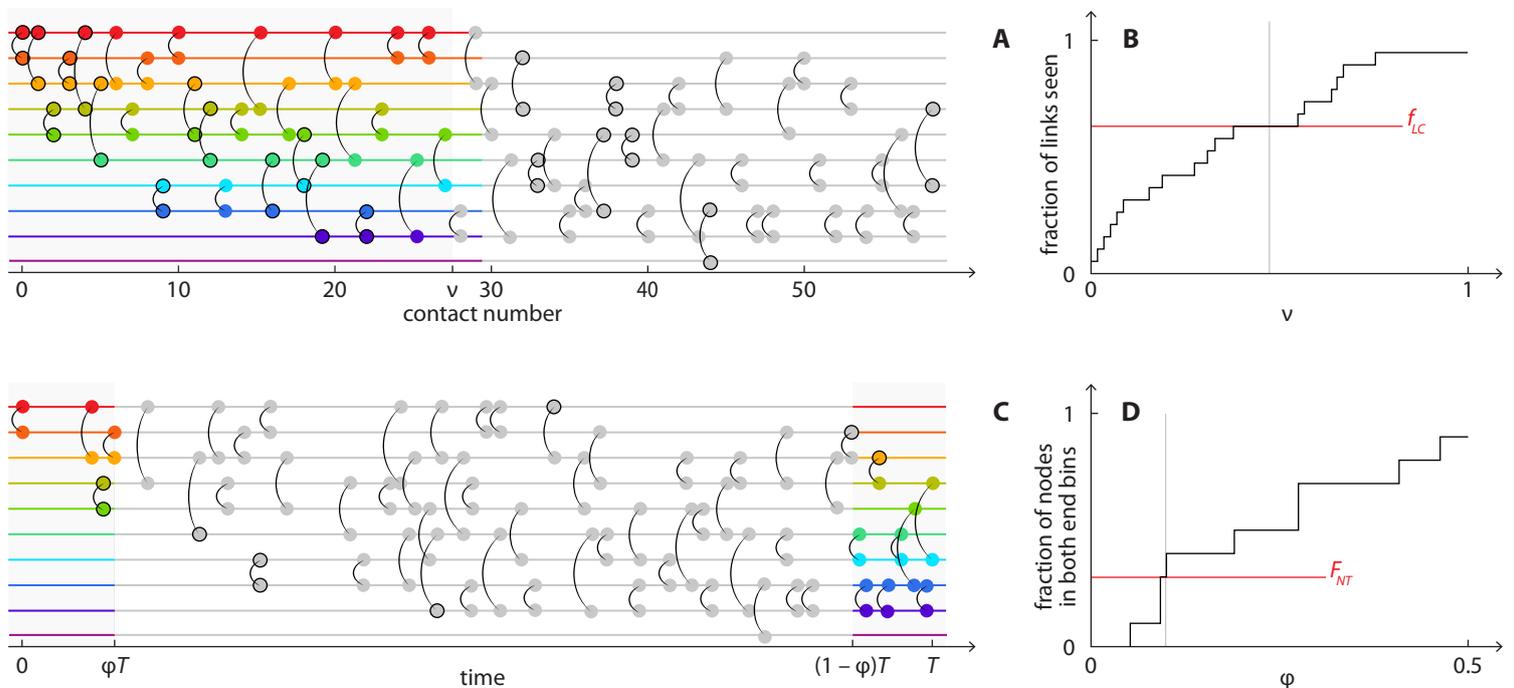

**Fig. 3. Illustration of two descriptors of temporal network structure, $f_{LC}$ and $F_{NT}$.** The measure illustrated in A and B, $f_{LC}$, uses the order of the contact to separate the contacts; the measure in C and D, $F_{NT}$, uses the real time. Panels A and C are time-line representations of a temporal network data set. Each horizontal line represents an individual. A contact between two individuals is indicated by a vertical arc. In A and B, we focus on the first contact between a pair of nodes. We measure the fraction of the number of node pairs that have been in direct contact when a fraction ν of the total number of contacts has been observed. This fraction is plotted against ν in B. The value at ν = 1/2 defines $f_{LC}$. In the timeline (A) we highlight the first half contacts, which contribute to the calculation of $f_{LC}$, in color and the first contact between each node pair by black contours. In panels C and D, we illustrate the calculation of $F_{NT}$, which looks at nodes (rather than links) present in both the first and last time interval of width φ (measured as a fraction of the sampling time), shown in color in the timeline (C). The fraction of such nodes as a function of φ is graphed in D. $F_{NT}$ is defined as the value at φ = 0.05.

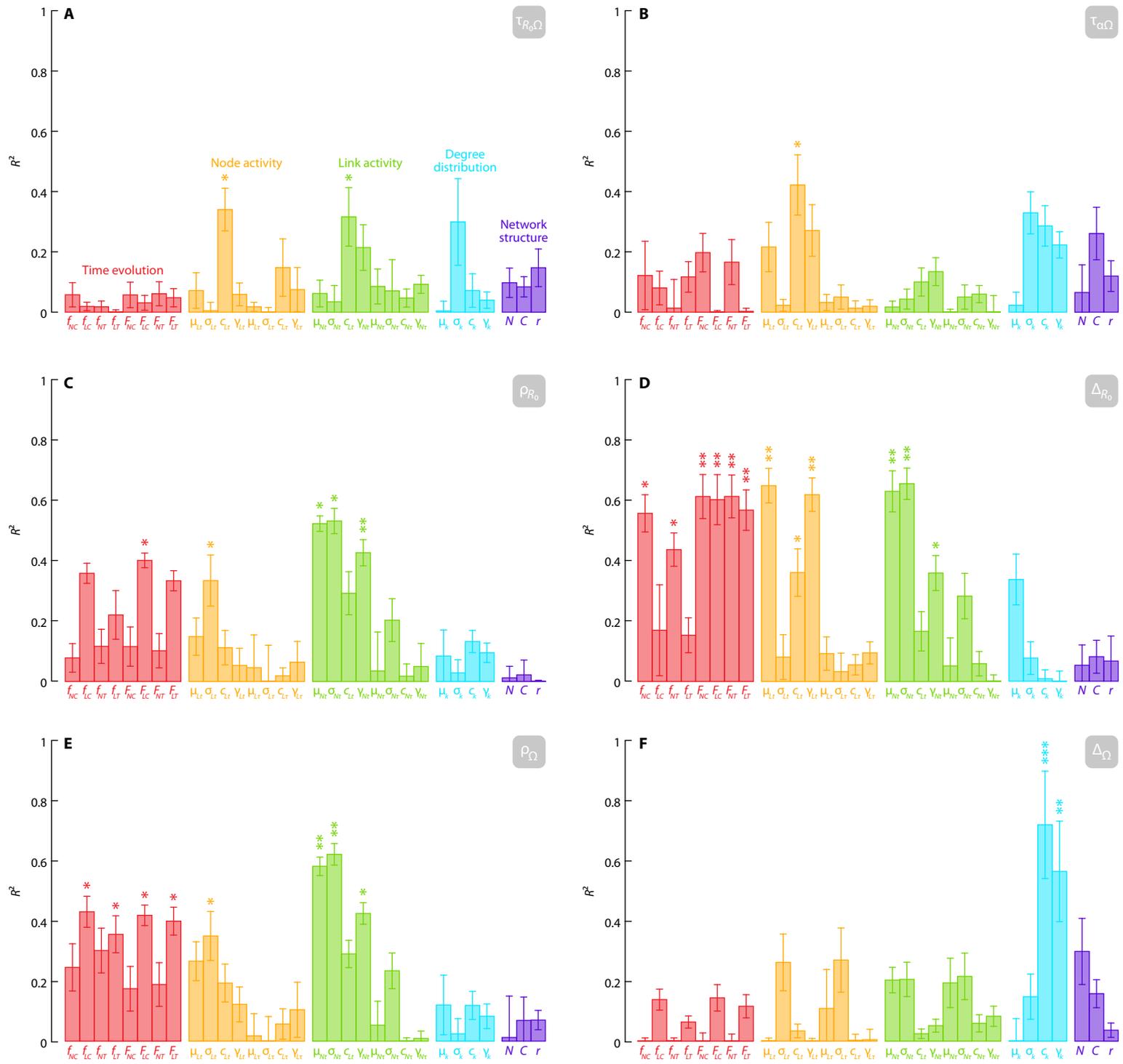

**Fig. 4. The coefficient of determination $R^2$ between the shape descriptors of the $R_0$ vs. $\Omega$ point cloud and network descriptors.** The error bars are standard errors estimated by the jackknife resampling method. *: $p < 0.05$, **: $p < 0.01$, ***: $p < 0.001$.